\documentclass[aps,prb,twocolumn,superscriptaddress,showpacs]{revtex4}
\usepackage{amsmath}
\usepackage{hyperref}
\usepackage{graphicx}
\begin{document}
\title{Tailoring magnetoresistance through rotating Ni particles}
\author{Steven Achilles}
  \email{steven.achilles@physik.uni-halle.de}
  \affiliation{Institute of Physics, Martin Luther University Halle-Wittenberg,  D-06099 Halle, Germany}
\author{Michael Czerner}
  \affiliation{Institute of Physics, Martin Luther University Halle-Wittenberg,  D-06099 Halle, Germany}
  \affiliation{I. Physikalisches Institut, Justus Liebig University, D-35392 Giessen, Germany}
\author{Ingrid Mertig}
  \affiliation{Institute of Physics, Martin Luther University Halle-Wittenberg,  D-06099 Halle, Germany}
\date{\today}

\begin{abstract}
We present \textit{ab initio} studies for different Ni nanocontacts and show changes in the conductance of such constrictions due to atomic rearrangements in the contact. In particular we consider a Ni particle and show that the magnetoresistance can change from a few to 50 \% and can even reverse sign as a function of the contact area formed between the particle and the leads.
\end{abstract}
\maketitle

\section{Introduction}

Nanocontacts in the limit of a few atoms represent the ultimate miniaturization of electronic devices and earned a lot of attention during the last years. In particular, ferromagnetic nanocontacts, their electronic structure and transport properties are of great interest and were intensively studied. 

The development  of different experimental techniques allows to fabricate metallic nanocontacts. In principle, the procedures can be classified into mechanical and chemical methods. The common mechanical techniques to produce contacts with a diameter in the order of a few nanometers are scanning tunneling microscopy (STM)\cite{Ohnishi98} and mechanically controllable break junctions (MCBJ)\cite{Scheer98,Agrait03}. It is known, that mechanical based methods are efficient and can be applied to a variety of materials. But they are not suitable for materials which tend to spontaneous breaking. Therefore electro-chemical methods like controlled electro deposition\cite{Li98,Li02,Elhoussine02,Calvo06} are preferable and allow the fabrication of contacts with a diameter of about $1~mm$ down to a few nanometers. The deposition of the ions is regulated by an applied voltage. The disadvantage is,  these methods are hard to control for nanocontacts in the limit of a few atoms\cite{Calvo06}.

Most experiments characterize the contact by measuring the conductance through the nanoconstriction. Monovalent metals like Cu\cite{Ohnishi98,Ludoph:2000p163,Agrait03}, Ag\cite{Ludoph:2000p163,Agrait03}, Au\cite{Ludoph:2000p163,Agrait03}, and Na\cite{Ludoph:2000p163,Agrait03} show integer values of the conductance quantum in the order of a few $g_0=2e^2/h$ down to $1g_0$. The conductance is very sensitive against tiny structural changes in the contact region which is reflected in the conductance histogram.

During the last years, ferromagnetic contacts consisting of Fe\cite{Untiedt04}, Co\cite{Rodrigues03,Untiedt04} and Ni\cite{Li02,Calvo06,Ono99,Garcia99,Garcia00,Garcia01,Elhoussine02,Chopra02,Chopra05,Untiedt04,Sullivan05,Viret02,Sirvent96} became very popular. Ferromagnetic contacts offer a further degree of freedom since the spin degeneracy is lifted. In particular, the relative orientation of the lead magnetization can be changed by external magnetic fields. As a result a domain wall is geometrically pinned in the constriction which changes the transport properties and can cause giant magnetoresistance (MR) expressed by the MR ratio
\begin{equation}\label{eqn:mr_ratio}
MR=\frac{g_{P}-g_{AP}}{g_{AP}}\times 100\%~.
\end{equation}
Here $g_P$ denotes the conductance for parallel alignment of the magnetic moments in the leads and $g_{AP}$ is the conductance for systems with anti-parallel lead magnetization.

The goal of the experiments was to reach large differences between $g_P$ and $g_{AP}$, i.e. large MR ratios. Among all the magnetic systems Ni was most of all analyzed \cite{Li02,Calvo06,Ono99,Garcia99,Garcia00,Garcia01,Elhoussine02,Chopra02,Chopra05,Untiedt04,Sullivan05,Viret02,Sirvent96}. Unfortunately, the results have been strongly contradictory also in comparison to theory\cite{Jacob05,Smogunov06,Pauly06}.

Some experimental results showed a half-integer conductance without an applied external magnetic field\cite{Elhoussine02,Sullivan05}, while other experiments did not support these observations\cite{Sirvent96,Viret02,Untiedt04}. Also in presence of magnetic fields the situation was confusing. Ono \textit{et al.}\cite{Ono99} reported a switching for the conductance from $2\,e^2/h$ to $e^2/h$ at low fields. Other authors claimed no changes by magnetic fields\cite{Untiedt04,Elhoussine02}. Furthermore, magnetostriction might be an additional source of uncertainty\cite{Egelhoff:2004p13094,Mallett:2004p13086}.

Some papers reported about extraordinary large MR ratios in Ni nanocontacts\cite{Garcia99,Garcia01,Chopra02,Sullivan05}, while others found maximum values of $40\,\%$\cite{Viret02}.

Theoretical investigations\cite{Jacob05,Smogunov06,Pauly06} did not confirm neither the half-integer conductance nor huge MR ratios.

The aim of this paper is to investigate structural changes in Ni nanocontacts and their effect on the conductance. Besides tensile and compressive strain we consider changes of the contact area between a nanoparticle and the leads which can be realized by a rotating particle. For this purpose we simulate a Ni particle consisting of four atoms situated at the corner of a tetrahedron in different positions with respect to the leads. For all configurations we assume collinear magnetic order and investigate the parallel and anti-parallel lead magnetization. A treatment of the non-collinear magnetic order is, in principle, possible but numerically very demanding for complex contact geometries\cite{Czerner:2008p782,Czerner:2008p784,Czerner:2010p12890}.

\section{Method}

Our electronic structure calculations are based on density functional theory (DFT)\cite{Hohenberg64} and their extension to spin-dependent DFT\cite{Barth72}. We use the Korringa-Kohn-Rostoker formalism (KKR)\cite{Korringa47,Kohn54,Zeller:1995p11296,Papanikolaou02} in the framework of the local spin density approximation (LSDA)\cite{Barth72} with the parameterization of the exchange-correlation potential introduced by Vosko, Wilk and Nusair\cite{Vosko80}.

For spherical potentials in the atomic sphere approximation and in cell-centered coordinates the Green's function writes as
\begin{equation}\label{eqn:GF_KKR}
\begin{split}
G(E;\mathbf{r}+\mathbf{R}^n,\mathbf{r}+\mathbf{R}^{n'})
& = \delta_{n,n'}\sqrt{E}\sum_{L} 
R_L^n(\mathbf{r}_<)\,H_L^{+\,n}(\mathbf{r}_>)\\
& + \sum_{L,L'}R_L^n(\mathbf{r})\,G^{n,n'}_{L,L'}(E)\,
R_{L'}^{n'}(\mathbf{r}')\,.
\end{split}
\end{equation}
In this notation, the angular momenta $(l,m)$ are represented by the index $L$ and $n$ labels the atomic sites. The wavefunctions $R_L^n(\mathbf{r})$ and $H_L^{+~n}(\mathbf{r})$ are the regular and irregular solutions of the single-site Kohn-Sham equations. The abbreviation $\mathbf{r}_>$ ($\mathbf{r}_<$) denotes the vector with larger (smaller) absolute value of $\mathbf{r}$ and $\mathbf{r}\,'$. In Eq.~\ref{eqn:GF_KKR}, $G^{n,n'}_{L,L'}(E)$ are the structure constants as a function of energy, atomic sites, and angular momenta.

The Green's function of the nanocontact is calculated in two steps. First, the Green's function of the host $\mathring{G}$ consisting of two semi-infinite leads separated by a vacuum spacer was calculated self-consistently. The structure of the leads was assumed to be ideal fcc with (001) surfaces.

Second, the Green's function of the suspended nanocontact $G$ was calculated by a Dyson equation:
\begin{equation}
\begin{split}
G^{n,n'}_{L,L'}(E)
& = \mathring{G}^{n,n'}_{L,L'}(E)\\
& + \sum_{n''L''} \mathring{G}^{n,n''}_{L,L''}(E)\,\Delta t^{n''}_{l''}(E) \,G^{n'',n'}_{L'',L'}(E)\,.
\end{split}
\end{equation}
The difference between the transition matrix  $t^n_l(E)$ of the real system and the matrix $\mathring{t}^n_l(E)$ of the host system for the single scattering potential on site $n$ is denoted by $\Delta t^{n}_{l}(E)$.

Baranger and Stone\cite{Baranger89} showed that the conductance $g$ at zero temperature can be calculated via
\begin{equation}\label{eqn:conductance_BS}
\begin{split}
g
& = \frac{e^2\hbar^3}{8 m^2 \pi}\times\\
& \int_{S_1}\int_{S_2}
d\mathbf{s_1}\cdot G^+(E_F;\mathbf{r}_1,\mathbf{r}_2)
\overleftrightarrow{\nabla}_{1}\!\!\overleftrightarrow{\nabla}_{2}
G^-(E_F;\mathbf{r}_2,\mathbf{r}_1)\cdot d\mathbf{s_2}\,,
\end{split}
\end{equation}
where $G^\pm(E_F;\mathbf{r}_1,\mathbf{r}_2)$ is the retarded (+) or advanced (-) Green's function at the Fermi level and $\overleftrightarrow{\nabla}_i$ is a double-sided derivative with respect to $\mathbf{r}_i$. The integrals in Eq.~\ref{eqn:conductance_BS} are taken over the surfaces $S_1$ and $S_2$ located in the left and right lead indicated by the surface elements $d\mathbf{s_1}$ and $d\mathbf{s_2}$ in the ideal leads, respectively.

Following Ref.~\onlinecite{Mavropoulos04}, the conductance can be written in terms of the structure constants $G^{n,n'}_{L,L'}(E)$ and the KKR current matrix elements
\begin{equation}
J_{L,L'}^n(E_F)=\frac{1}{\Delta}\int_{V} d^3r~R_L^n(E_F;\mathbf{r})\nabla R_{L'}^n(E_F;\mathbf{r})^*\,,
\end{equation}
as
\begin{equation}
\begin{split}
g & = \frac{e^2\hbar^3}{8 m^2 \pi}\times\\
  &   \sum_{n,n'}\sum_{L,L'}\sum_{L'',L'''}(J_{L,L''}^n(E_F)-J_{L'',L}^{n\,*}(E_F))\times\\
  &     (J_{L',L'''}^{n'}(E_F)-J_{L''',L'}^{n'\,*}(E_F)) G^{n,n'}_{L,L'}(E_F)G^{n,n'\,*}_{L'',L'''}(E_F)\,.
\end{split}
\end{equation}
The summation over $n$ and $n'$ is restricted to atomic spheres in the planes $S_1$ and $S_2$. According to B{\"u}ttiker\cite{Buttiker:1988p3739}, the total conductance $g$ can be understood as a superposition of individual transmission $T_m^{\sigma}$ probabilities through individual eigenchannels $m$ of spin $\sigma$
\begin{equation}
g=\frac{e^2}{h}\sum_{\sigma,m} T_m^{\sigma}\,.
\end{equation}
 
Scheer \textit{et al.}\cite{Scheer98} extracted the individual transmission probabilities for the eigenchannels from current-voltage-characteristics of atomic sized contacts between superconducting leads. In particular, a projection of the conductance onto the atomic orbitals at the central atom in the contact is used\cite{Cuevas98}. Bagrets \textit{et al.}\cite{Bagrets07} developed a method for the decomposition of the conductance into eigenchannels based on a KKR-basis set depending on site and angular momenta which is applied in this paper.

In the general case of propagating Bloch functions the conductance channels are classified by the point group symmetry of the nanocontact. In case of $C_{4V}$ symmetry the eigenchannels are characterized by the irreducible representations $\Delta_1$,$\Delta_2$, $\Delta_{2'}$, and $\Delta_5$. A projection of the irreducible representations onto angular momenta up to $l_{max}\leq 3$ is given in Tab.~\ref{tab:spherical_harmonics_symmetry}.

\begin{table}
\caption{\label{tab:spherical_harmonics_symmetry} Irreducible representations of the group $C_{4V}$ and the corresponding angular momentum characters up to $l_{max}\leq 3$.}
\centering
\begin{ruledtabular}
\begin{tabular}{cl}
irreducible representation & angular momentum character \\
\hline
\hline
$\Delta_1$	& $s$, $p_z$, $d_{3z^2-r^2}$ \\
\hline
$\Delta_2$	& $d_{x^2-y^2}$ \\
\hline
$\Delta_{2'}$	& $d_{xy}$ \\
\hline
$\Delta_5$	& $p_x$, $d_{xz}$ \\
\hline
$\Delta_5$	& $p_y$, $d_{yz}$ \\
\end{tabular}
\end{ruledtabular}
\end{table}

\section{Contact geometries}\label{sec:contacts}

\begin{figure}
  \includegraphics[scale=.35]{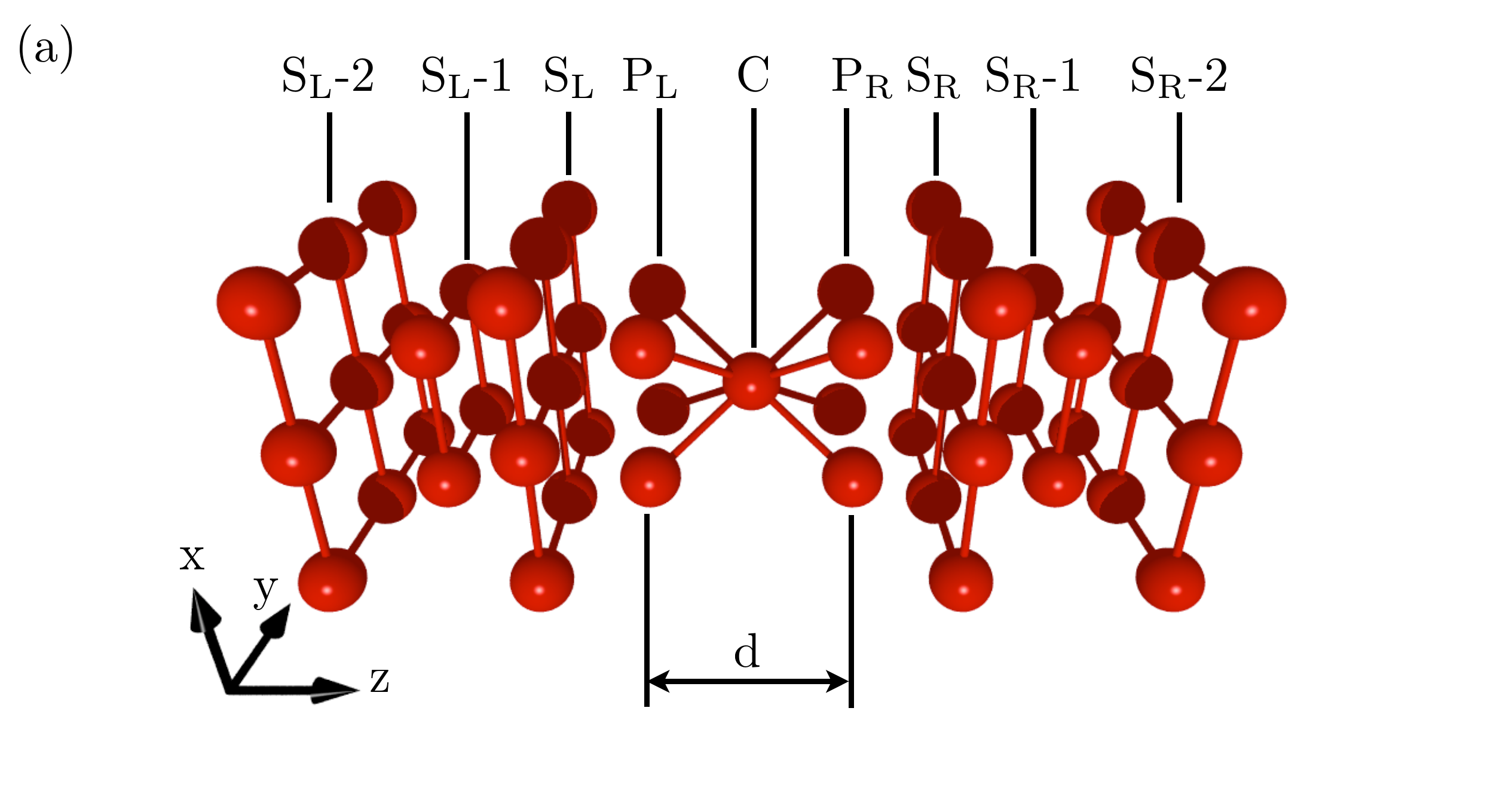}
  \includegraphics[scale=.23]{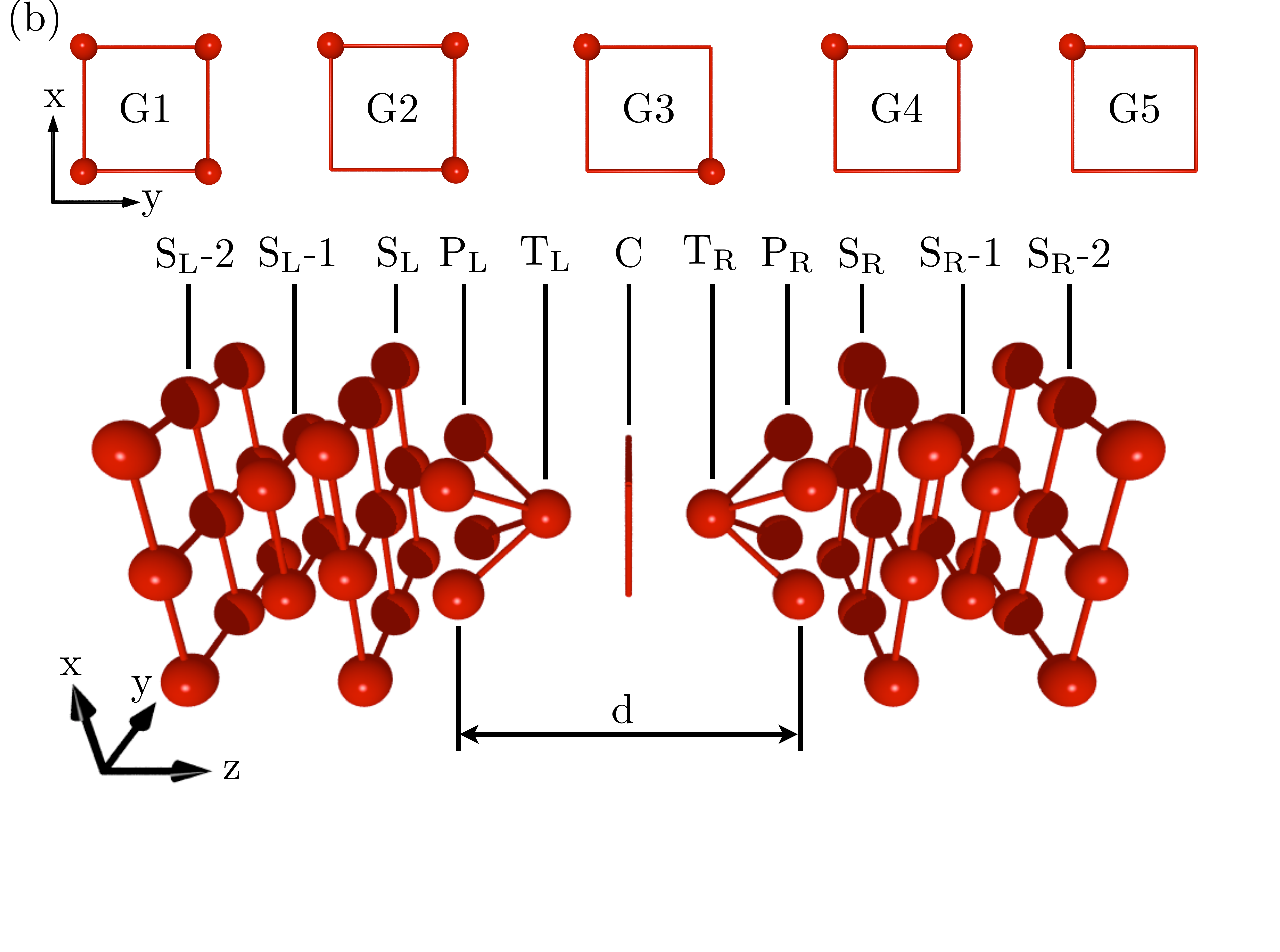}
  \includegraphics[scale=.23]{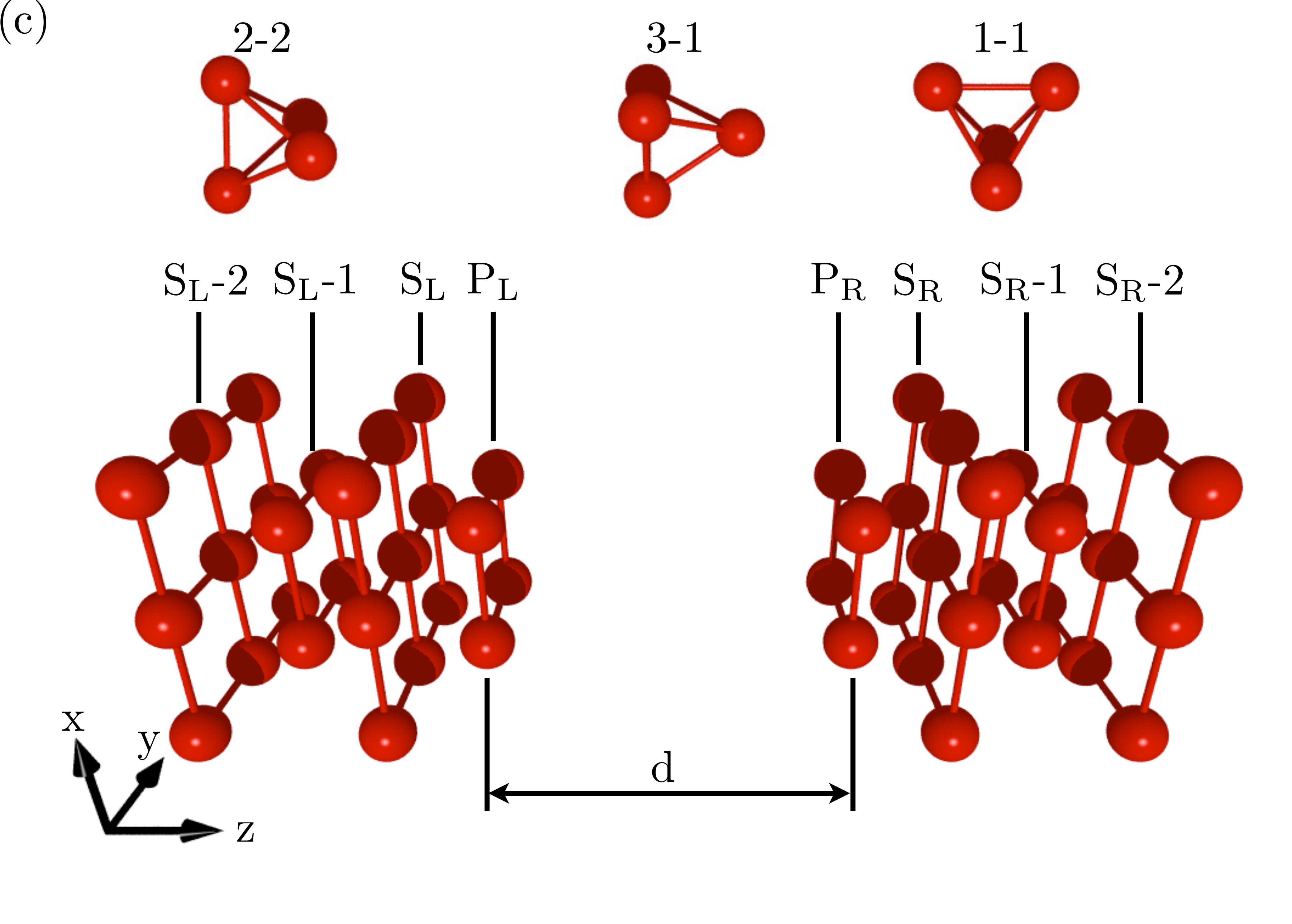}
  \caption{Sketch of the different nanocontacts suspended between macroscopic leads.\label{gfx:AC_geometric}}
\end{figure}

For our calculations, we consider three different kinds of Ni nanocontacts: a  single atomic contact (Fig.~\ref{gfx:AC_geometric}a), a few atom constriction (Fig.~\ref{gfx:AC_geometric}b), and a small Ni particle bridging two semi-infinite leads (Fig.~\ref{gfx:AC_geometric}c). For the single-atom contact and the few atom contact we focus on changes in the conductance caused by strain and symmetry separately. For the Ni particle both effects are superimposed.

All nanocontacts are sandwiched between two Ni leads in fcc structure with an experimental lattice constant of $d_0=3.52$ \AA. The surface normal is (001). The first two contacts are suspended between two tips consisting of five atoms continuing the fcc lattice.

For the single-atom contact (Fig.~\ref{gfx:AC_geometric}a) the two tips are merged resulting in a single contact atom between four-atom plateaus ($P_L$, $P_R$). To simulate strain the distance $d$ between the plateaus is changed gradually while the single atom ($C$) is kept fixed in the middle.

To study the influence of symmetry changes in the contact region we embedded a square $(C)$ with Ni atoms at the corners in between the two tips (Fig.~\ref{gfx:AC_geometric}b). The corners of the square ($C$) and the tips ($T_{\mathrm{L}}$, $T_{\mathrm{R}}$) form a structure which is similar to fcc in (001) orientation. The number of Ni atoms on the central square ($C$) is gradually reduced from 4 to 1 resulting in the configurations G1 - G5 (see Fig.~\ref{gfx:AC_geometric}b).

Finally, we suspended a Ni particle between the plateaus (see Fig.~\ref{gfx:AC_geometric}c). The Ni particle consists of four atoms sitting at the corners of a regular tetrahedron. The edge length is equal to the next-nearest-neighbor distance of our fcc Ni. The center of gravity of the Ni particle is fixed in the middle between the plateaus and the distance between the plateaus is fixed at $d$=7.48\,\AA{}. The tetrahedron is assumed to be a freely rotating particle. We study three different configurations: a two-by-two (2-2), a one-by-one (1-1) and a three-by-one (3-1) configuration. Because of the fixed center of gravity of the particle, the distances between the plateaus ($P_L$, $P_R$) and the outermost atoms of the particle are changing as a function of the particular orientation of the tetrahedron. For simplicity we show in Tab.~\ref{tab:bonding_distances_particle} the corresponding spacings in (001)-direction ($d_{\bot,L}$, $d_{\bot,R}$) between the left or right plateau and the outermost atoms on the left or right hand side of the particle, respectively.

\begin{table}
\caption{\label{tab:bonding_distances_particle} Distances between the plateaus ($P_L$, $P_R$) and the outermost atoms on the left or right hand side in (001)-direction of the Ni particle for the (2-2), (1-1) and (3-1) configuration.}
\centering
\begin{ruledtabular}
\begin{tabular}{c|cc}
configuration & distance $d_{\bot,L}$ (\AA{}) & distance $d_{\bot,R}$ (\AA{}) \\
\hline
(2-2) & 2.85 & 2.85 \\
(3-1) & 3.23 & 2.21 \\
(1-1) & 2.41 & 2.41 \\
\end{tabular}
\end{ruledtabular}
\end{table}

\section{Results}

\subsection{Single atom contact and strain}\label{sec:SAC}

To study the influence of strain we use the single-atom contact (Fig.~\ref{gfx:AC_geometric}a). The ideal fcc lattice with the bulk lattice constant $d_0$ is the ''reference'' system.

\begin{figure}
  \includegraphics[scale=0.65]{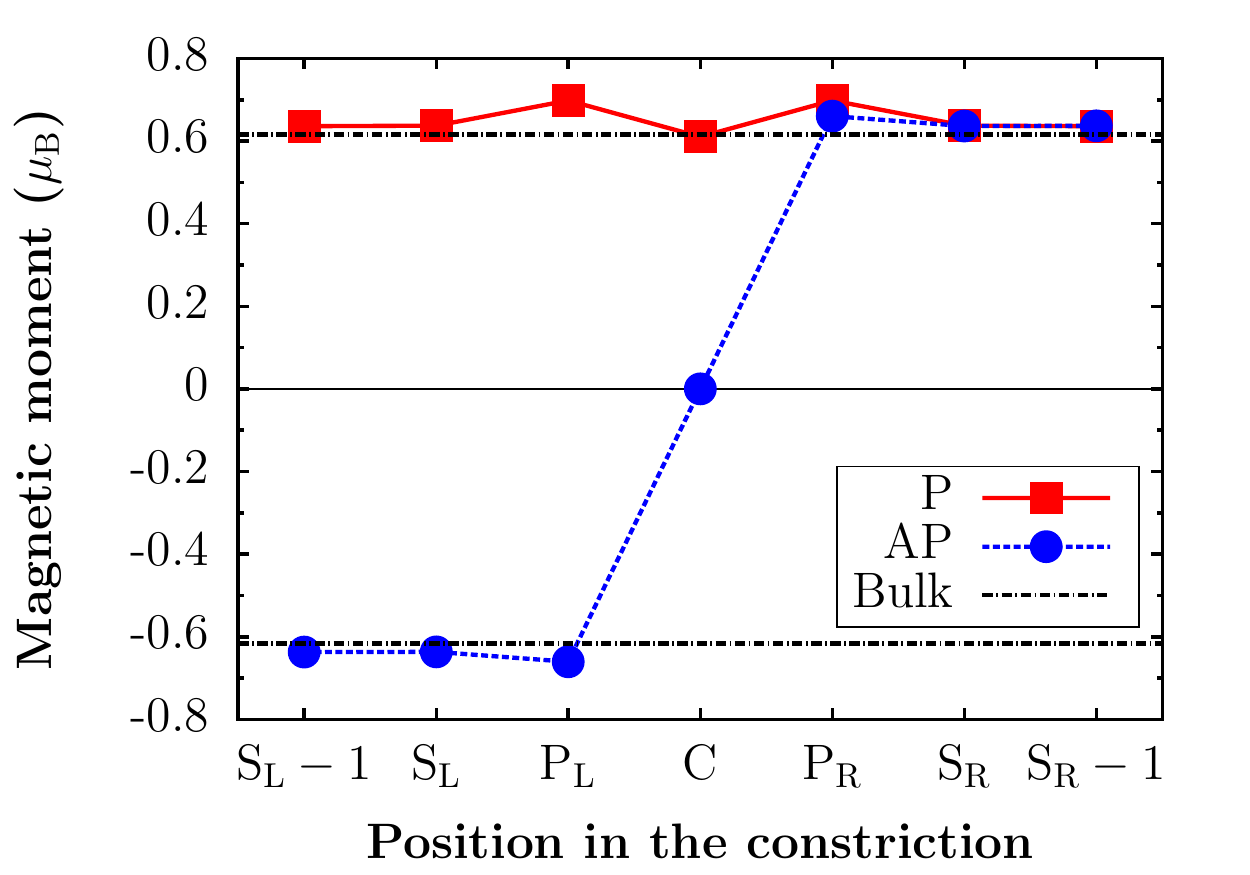}
    \caption{(color online) Magnetic moments inside the constriction for the single-atom contact in ideal bulk structure for parallel (P) and anti-parallel (AP) lead magnetization in comparison to the bulk magnetic moment.\label{gfx:SAC_magnetic}}
\end{figure}

Furthermore, we define a relative length variation of the plateau-plateau distance as $\Delta d / d_0$, with $d_0$ the bulk lattice constant and $\Delta d=d-d_0$, the deviation of the plateau-plateau distance from the ''reference'' system. The relative length variation between the plateaus was changed in the range of $-10\%$ up to $+10\%$ in steps of $5\%$.

To characterize the single-atom contact magnetically the magnetic moments for parallel and anti-parallel alignment of the lead magnetization are shown in Fig.~\ref{gfx:SAC_magnetic}. For the parallel alignment (P) , the distribution of the magnetic moments is symmetric with respect to the central atom in the single-atom contact. The magnetic moments in the planes $S_L$-1 ($S_R$-1) and $S_L$ ($S_R$) are very similar to the bulk moment $m_{Bulk}=0.62~\mu_B$. The magnetic moment on top of the surface $P_L$ ($P_R$) is increased with respect to the reduced coordination number. The magnetic moment at the central atom is again bulk-like.

For the anti-parallel configuration (AP) a domain wall is pinned in the constriction. As a consequence based on a collinear treatment of magnetism the magnetic moment at the central atom is suppressed (Fig.~\ref{gfx:SAC_magnetic}) because of frustration.

\begin{figure}
  \includegraphics[scale=.55]{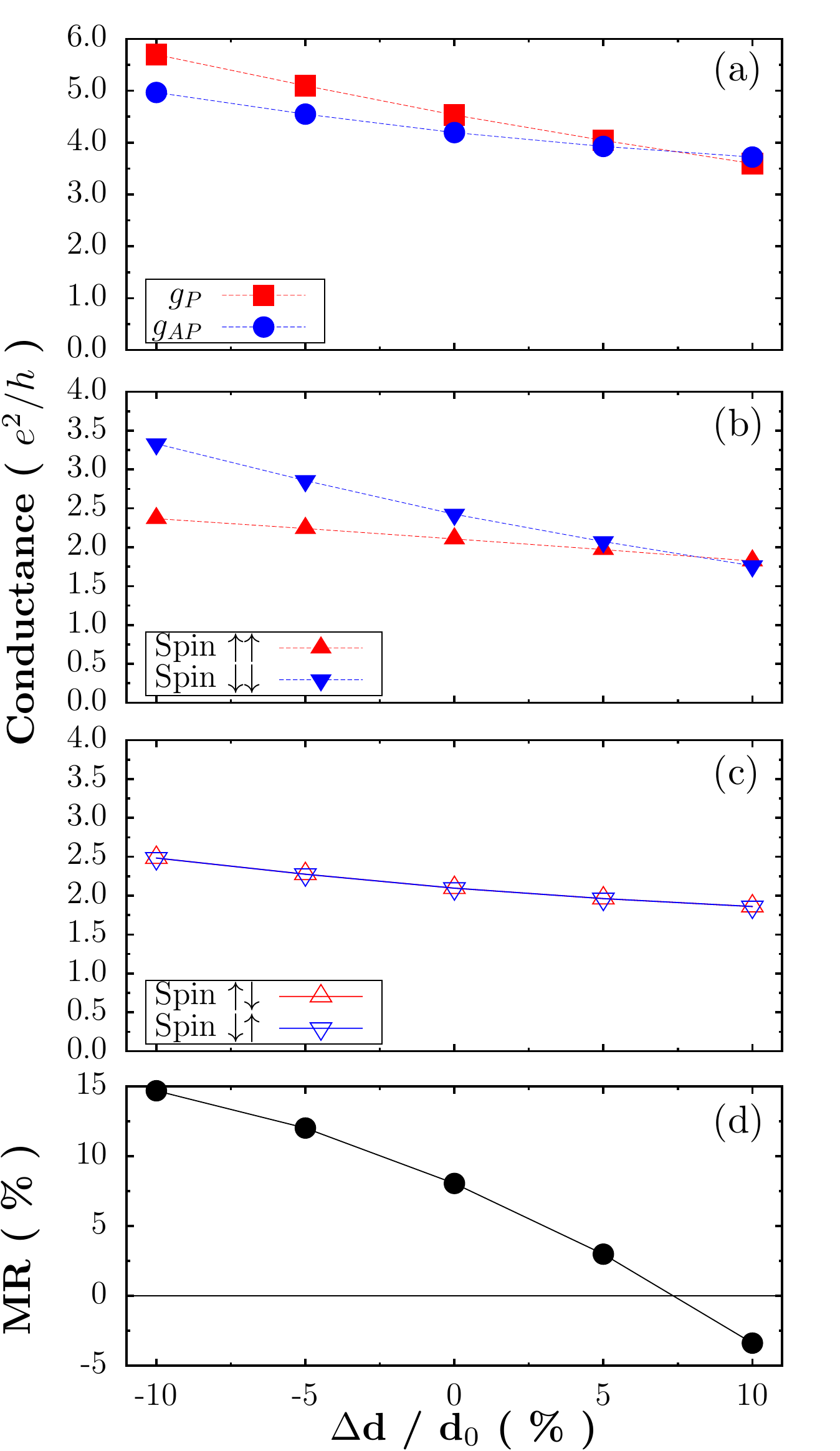}
    \caption{(color online) Total conductance for (a) parallel ($g_P$) and anti-parallel ($g_{AP}$) lead magnetizations of the single-atom contact, spin-resolved conductance for (b) parallel and (c) anti-parallel lead magnetizations, and (d) MR ratio as a function of the relative length variation.\label{gfx:SAC_g_all}}
\end{figure}

The total conductance values of the single-atom contact as a function of strain (Fig.~\ref{gfx:SAC_g_all}a) are in good agreement with experiments\cite{Sirvent96,Li02,Viret02,Agrait03,Untiedt04}. Tensile strain reduces the conductance whereas compressive strain increases the conductance. This behavior is intuitively clear since the overlap of the atomic orbitals is reduced or increased, respectively. The overlap of the wavefunction is in a simple tight-binding-like picture a measure for the ''hopping'' of electrons. The change of the conductance for P and AP configuration with strain differs. An explanation of the different decay rates will be given below in terms of eigenchannels.

The spin-resolved conductance for P and AP configuration is shown in Fig.~\ref{gfx:SAC_g_all}b and c, respectively. The current of majority electrons changes slightly with strain whereas the current of minority electrons strongly reflects strain in the P configuration. This fact will be discussed in terms of eigenchannels.

The spin-resolved conductance of the AP configuration is degenerate with respect to spin because of symmetry ($g_{\uparrow\downarrow}=g_{\downarrow\uparrow}$). The spin-resolved conductance is nearly the same as $g_{\uparrow\uparrow}$.

The MR ratio (Eq. \ref{eqn:mr_ratio}) is plotted in Fig.~\ref{gfx:SAC_g_all}d. The MR ratio is positive in the range of $-10~\%$ to $\sim +7.5~\%$ relative length variation. The sign reversal corresponds to a crossing of the conductance curves for the parallel and the anti-parallel configuration. For a relative length variation of $10~\%$, the conductance value for the anti-parallel configuration is higher than in the parallel case which causes a negative MR ratio. The maximum value of $14.7~\%$ corresponds to a relative length variation of $-10\%$. The positive MR ratios are qualitatively in agreement with experiment\cite{Viret02}.

\begin{figure*}
  \includegraphics[scale=.41]{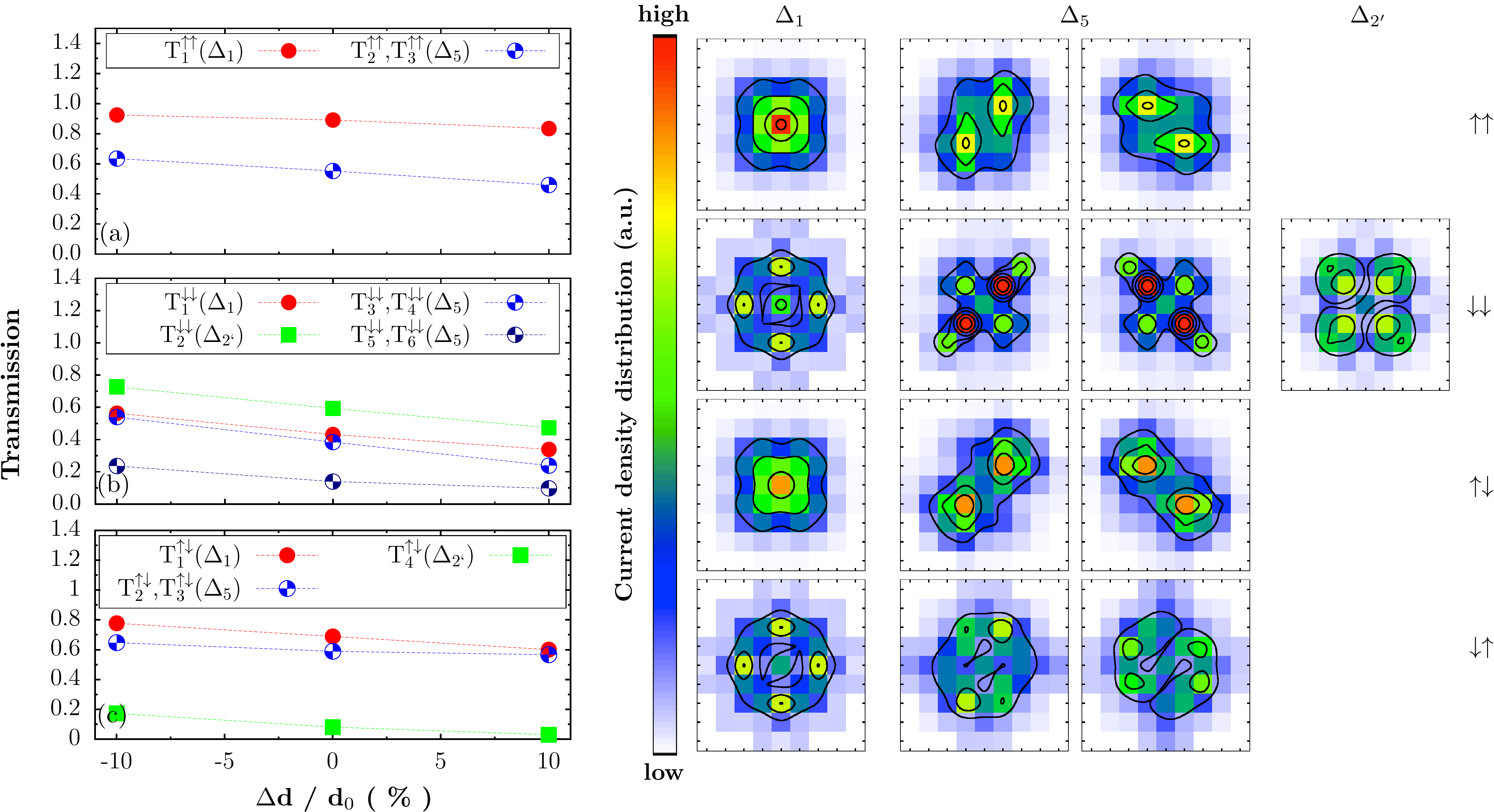}
  \caption{(color online) \textit{(left)} Symmetry resolved transmission probabilities for (a) majority and (b) minority states for the P configuration, and  (c) transmission probabilities for the AP configuration for the single-atom contact as a function of the relative length variation. \textit{(right)} Current density distribution in the plane $S_L$-2 for a relative length variation of $\Delta d/d_0=0\,\%$. \label{gfx:SAC_T_all} }
\end{figure*}

The investigated nanocontact belongs to the point group $C_{4V}$. The results of the eigenchannel decomposition as a function of strain are given in Fig.~\ref{gfx:SAC_T_all}, majority and minority channels in the P configuration in Fig.~\ref{gfx:SAC_T_all}a and b, channels of the AP configuration in Fig.~\ref{gfx:SAC_T_all}c.

Ni is a ferromagnetic material with the configuration $4s^2\,3d^8$ for the conduction electrons, which leads  to $2+8=10$ (partially) open channels for an ideal one-dimensional nanowire. The single atom contact is not a one-dimensional structure and therefore even more than 10 conduction channels could exist. The following discussion is, however, restricted to channels with transmission probabilities $T_m^\sigma > 0.10$.

In the P configuration, three majority channels are partially open (Fig.~\ref{gfx:SAC_T_all}a). The main contribution corresponds to the irreducible representation $\Delta_1$. If we would analyze the channel compositioon, we would find $s$, $p_z$, and $d_{3z^2-r^2}$ orbitals at the single-atom bottleneck of the contact (see Tab.~\ref{tab:spherical_harmonics_symmetry}). These particular orbitals are oriented in $(001)$-direction, i.e. the current direction. The other two channels are compatible with the representation $\Delta_5$ and consist approximately of atomic $p_x$ $(p_y)$ and the $d_{xz}$ $(d_{yz})$ orbitals at single-atom position ($C$), or linear combinations of them. These orbitals are oriented perpendicular to the current direction. Therefore, small changes in the distance $d$ cause strong changes in the overlap of the wavefunctions. The reduced overlap induces a reduced conductance. This is the reason for the slightly enhanced decay of the channels T$_{2}^{\uparrow\uparrow}$ and T$_{3}^{\uparrow\uparrow}$ in comparison with T$_1^{\uparrow\uparrow}$. On the right hand side of Fig.~\ref{gfx:SAC_T_all}, the current density contributions for different eigenchannels in the plane $S_L$-2 are shown. From the symmetry of the current distribution one can see, that the channel T$_1^{\uparrow\uparrow}$ is dominated by $s$ and $p_z$ contributions, and the channel T$_{2}^{\uparrow\uparrow}$ (T$_{3}^{\uparrow\uparrow}$) is dominated by $p_x$ ($p_y$) orbitals.

For the minority electrons (see Fig.~\ref{gfx:SAC_T_all}b), six channels are partially open. Their corresponding irreducible representations are $\Delta_1$, $\Delta_5$, and $\Delta_{2'}$. The comparison of the current density distributions shows the strong influence of $d$ electrons which is already expected from the local density of states at the Fermi level (not shown). The strong $d$ character at the bottleneck results in a strong distance dependence, which is illustrated in Fig.~\ref{gfx:SAC_T_all}b. For all channels with representations $\Delta_5$ and $\Delta_{2'}$ the decay of conductance is slightly enhanced by increasing strain compared to channels with $\Delta_1$ representation. The number of conduction channels for the minority electrons is larger than for the majority states. The decay rates of the channel contributions for minority states are very similar to the decay rates of majority electrons. But the number of conduction channels for minority states is larger than for majority states which is the origin of the stronger decay of the minority conductance in P configuration with increasing strain (Fig.~\ref{gfx:SAC_g_all}b).

For the anti-parallel case, the analysis of conduction eigenchannels has been performed, too. The resulting transmission probabilities are plotted in Fig.~\ref{gfx:SAC_T_all}c. Here, four channels per spin with the corresponding representations $\Delta_1$, $\Delta_5$, and $\Delta_{2'}$ are formed. The channel T$_4^{\uparrow\downarrow}$ with representation $\Delta_{2'}$ is marginally open. For higher relative length variations, this channel is closed.

A comparison of the conduction channels shows that the anti-parallel alignment of lead magnetization changes the composition of the channels and their transmission probabilities drastically. The relatively strong reduction of the conductance in the parallel case is mainly forced by $d$-like contributions in the minority spin, while the majority contributions change only slightly. The channels in the anti-parallel case show the same behavior like the majority states in the P configuration. In particular, the decay rates of the individual transmission probabilities, the number of open channels and their contributions to the transmission are very similar.

Finally, the larger decay of conductance in the minority spin compared with the majority contributions in the P configuration favors the crossing (see Fig.~\ref{gfx:SAC_T_all}a, b and Fig.~\ref{gfx:SAC_g_all}b). The spin-resolved conductances, their magnitude as well as the number of conduction channels for the anti-parallel alignment of lead magnetization are very similar to the majority contributions in the P configuration (cf. Fig.~\ref{gfx:SAC_T_all}a and c). Because of degeneracy in the AP configuration no crossing of the spin-resolved conductances is observed. The crossing of the total conductances for P and AP configuration (see Fig.~\ref{gfx:SAC_g_all}a) and the corresponding change in the sign of the MR ratio (see Fig.~\ref{gfx:SAC_g_all}d) is caused by an imbalance in the number of contributing channels to the spin-resolved conductances for both magnetic configurations.

\subsection{Small Ni clusters}

In a next step we focus on the influence on symmetry on the transmission and therefore we replace the single-atom contact by a few atom constriction.

Details of the contact geometry have been discussed in Sec.~\ref{sec:contacts}. The symmetry of the contact is gradually reduced from $C_{4V}$ to lower symmetry by reducing the number of atoms in the central plane (see Fig.~\ref{gfx:AC_geometric}b). Except for G3 where in comparison to G2 one additional rotation by $180^\circ$ and two reflections with respect the two atoms in $C$ leaves the contact invariant.

Before we discuss the conductance as a function of contact geometry we will concentrate on the magnetic structure. In the case of parallel lead magnetization the moments are ferromagnetically ordered in the ground state. The magnetization profile is symmetric with respect to the central plane $C$. As a consequence of a reduced coordination numbers the magnetic moments increase. This behavior can be demonstrated by considering the magnetic moments at different positions in the contact. The moments in the layers $S_L$-1 and $S_R$-1 (see Fig \ref{gfx:AC_geometric}b) are a slightly enhanced ($m=0.63\,\mu_B$) compared to the bulk value ($m_{Bulk}=0.62\,\mu_B$). The magnetic moments in the surface layers ($S_L$ and $S_R$) have a value of $m=0.64\,\mu_B$. A further reduction of the coordination number causes enhanced magnetic moments inside the contact ($P_L$, $T_L$, $C$, $T_R$, and $P_R$) up to a maximum value of $m=0.80\,\mu_B$.

The case of anti-parallel orientation of the lead magnetization is more complicated. We are looking for the lowest energy configuration under the constraint of the anti-parallel orientation of the lead magnetization and the assumption of collinear magnetic order. As a result an abrupt domain wall is pinned between left and right electrode. Except geometry G4, the domain wall is located between the central atoms ($C$) and the left or right tip atom ($T_L$, $T_R$). For symmetry reasons both positions are equivalent. The domain wall of G4 is located at the central plane. The moments of the two atoms on the central plane are suppressed. The moments in and near the surface are higher than the bulk values. The magnetic moments on the central plane are in the order of $m\approx 0.7\mu_B$.

\begin{table}
\caption{\label{tab:SAC_FAC_g} Comparison of the spin resolved conductance values of the unconstrained P configuration of the sinlge-atom contact with the G1 configuration.}
\centering
\begin{ruledtabular}
\begin{tabular}{c|c|c}
conductance & single-atom contact & few atom contact \\
\hline
$g_{\uparrow\uparrow}$ [$e^2/h$]		& 2.10	& 2.20 \\
$g_{\downarrow\downarrow}$ [$e^2/h$]	& 2.43 	& 0.97 \\
\end{tabular}
\end{ruledtabular}

\end{table}
\begin{table}
\caption{\label{tab:SAC_FAC_channels} Spin-resolved transmission probabilities for the dominating eigenchannels of the unconstrained P configuration of the single-atom contact with the G1 configuration.}
\centering
\begin{ruledtabular}
\begin{tabular}{r|c|c}
\multicolumn{1}{c|}{channels} 	& \multicolumn{2}{c}{transmission for} \\
 & single atomic contact & few atom contact\\
\hline
$T_1^{\uparrow\uparrow} (\Delta_1)$	& 0.88	& 0.71 \\
$T_2^{\uparrow\uparrow},T_3^{\uparrow\uparrow} (\Delta_5)$	& 0.54	& 0.74 \\
\hline
$T_1^{\downarrow\downarrow} (\Delta_1)$	& 0.38	& 0.12 \\
$T_2^{\downarrow\downarrow} (\Delta_{2'})$	& 0.51	& 0.13 \\
$T_3^{\downarrow\downarrow},T_4^{\downarrow\downarrow} (\Delta_5)$	& 0.50	& 0.29 \\
\end{tabular}
\end{ruledtabular}
\end{table}

\begin{table*}
\caption{\label{tab:FAC_g_MR} Conductance values and MR ratios for Ni nanocontacts consisting of small clusters in dependence on the geometrical configuration in the contact region.}
\centering
\begin{ruledtabular}
\begin{tabular}{c|ccc|ccc|c}
geometric	      	& \multicolumn{3}{c|}{parallel configuration} & \multicolumn{3}{c|}{anti-parallel configuration} & MR \\
configuration 	& $g_{\uparrow\uparrow}$ [$e^2/h$] 	& $g_{\downarrow\downarrow}$ [$e^2/h$]	& $g_{P}$ [$e^2/h$]
			& $g_{\uparrow\downarrow}$ [$e^2/h$]	& $g_{\downarrow\uparrow}$ [$e^2/h$]		& $g_{AP}$ [$e^2/h$]
& [\%]\\
\hline
G1	& 2.20 & 0.97 & 3.18 & 1.18 & 0.78 & 1.97 & 61.38 \\
G2	& 1.28 & 0.98 & 2.27 & 1.45 & 0.99 & 2.44 & -7.25 \\
G3	& 1.23 & 0.91 & 2.14 & 1.04 & 0.71 & 1.75 & 22.22 \\
G4	& 0.79 & 0.73 & 1.52 & 0.57 & 0.57 & 1.15 & 32.28 \\
G5	& 0.92 & 0.69 & 1.61 & 0.57 & 0.69 & 1.26 & 28.21 \\
\end{tabular}
\end{ruledtabular}
\end{table*}

The conductance values and the corresponding MR ratios as a function of contact geometry are given in Tab.~\ref{tab:FAC_g_MR}. Two aspects determine the general trend. First, the larger the contact area the higher is the conductance. Second, the lower the symmetry of the constriction the lower is the conductance.

Before we discuss the effect of symmetry reduction on the conductance of the few-atom contact let us compare $g_P$ of the unstrained single-atom contact with the few-atom contact G1. Both contacts have the same symmetry $C_{4V}$. The conductance of the single-atom contact is however higher than the conductance of G1. The spin-resolved conductance values are compared for both contacts in Tab.~\ref{tab:SAC_FAC_g}. The majority contribution $g_{\uparrow\uparrow}$ is nearly unchanged whereas the minority contribution is drastically reduced which is caused by blocking of eigenchannels. From the eigenchannel analysis in Sec.~\ref{sec:SAC} we know that the minority contributions are mainly $d$-like and the majority contributions are $s$-$p$-like. In Tab.~\ref{tab:SAC_FAC_channels} the transmission probabilities of both configurations are compared.

The dominating channels of the majority electrons have $\Delta_1$ and $\Delta_5$ character. As discussed before $\Delta_5$ is twofold degenerate. The transmission probabilities of both configurations have the same order of magnitude.

The channels of the minority electrons have also $\Delta_1$, $\Delta_5$ and $\Delta_{2'}$ symmetry. The magnitude of the transmission probabilities is however reduced by more than a factor of two for the few-atom contact. The $d$ contributions are blocked in the geometry G1 since they have to pass two single-atom contacts successively which acts like a filter for the strongly localized $d$ states.

With the help of the eigenchannel decomposition we can explain the changes of $g_P$, in particular $g_{\uparrow\uparrow}$,  under the reduction of symmetry. Going from G1 to G2 by removing one atom the $C_{4V}$ symmetry is destroyed. The invariance under rotation around the $z$ axis is lost. The twofold $\Delta_5$ does not any more exist. The corresponding channels are blocked and the conductance is reduced by $1g_0$ (see Tab.~\ref{tab:FAC_g_MR}). Further reduction of the symmetry can be discussed analogously going from G2 to G4 or G5. In contrast, the conductance for G2 is higher than the conductance for G3, although the symmetry is increased going from G2 to G3. In this particular case the reduced contact area formed by the number of central atoms determines the conductance.

The changes of $g_{\downarrow\downarrow}$ under symmetry reduction are not that pronounced since the channels are already only partially open.

In the anti-parallel configuration the inversion symmetry with respect to the central plane is lost because of the magnetic order. As a result the $g_{AP}$ values are usually smaller than the $g_P$ ones. Incoming majority electrons are scattered into minority states and vice versa. New eigenchannels are formed.

A further reduction of symmetry in the central plane is not as important as in the parallel configuration since the pinned domain wall which is situated nearby the central plane acts already as a filter and requires a node of the eigenchannel. In some cases this fact can even cause $g_{AP}>g_P$ (see G2).

The individual conductance values are reduced for the few-atom contact with respect to the single-atom contact. The MR ratios are of the same order of magnitude in agreement with experiment\cite{Viret02}, can become even higher or reverse sign. Extraordinary MR ratios as measured\cite{Garcia99,Garcia01,Chopra02,Sullivan05} are however not obtained.

\subsection{Ni particle (tetrahedron)}
\begin{table*}
\caption{\label{tab:Particle_g} Spin-resolved and total conductance for different orientations of the Ni particle in parallel and anti-parallel configuration.}
\centering
\begin{ruledtabular}
\begin{tabular}{c|ccc|ccc}
orientation of the	& \multicolumn{3}{c|}{parallel configuration}	& \multicolumn{3}{c}{anti-parallel configuration} \\
tetrahedron		& $g_{\uparrow\uparrow}$ [$e^2/h$]			& $g_{\downarrow\downarrow}$ [$e^2/h$]	& $g_{P}$ [$e^2/h$]
				& $g_{\uparrow\downarrow}$ [$e^2/h$]		& $g_{\downarrow\uparrow}$ [$e^2/h$]		& $g_{AP}$ [$e^2/h$] \\
\hline
2-2		& 0.93 & 1.33 & 2.25 & 1.05 & 1.05 & 2.10 \\
3-1		& 0.90 & 0.98 & 1.89 & 0.81 & 0.70 & 1.51 \\
1-1		& 0.95 & 1.25 & 2.19 & 0.76 & 0.76 & 1.51 \\
\end{tabular}
\end{ruledtabular}
\end{table*}
In this paragraph we discuss the results for the Ni particle (see Fig.~\ref{gfx:AC_geometric}c).

From some experiments, it is known, that structural changes can affect the transport properties strongly\cite{Sirvent96,Agrait03,Untiedt04}. Especially, for nanoconstrictions formed by controlled electro deposition of ions, the geometrical configuration is still unclear \cite{Calvo06} and applied external fields, in principle, could cause atomic rearrangements inside the contact.

We assume a Ni particle that, in principle, can rotate freely between the leads caused by an applied bias or by interaction of the particle magnetization with an external magnetic field. In our investigations, we assume that the particle rotates with respect to a fixed center of gravity in the middle of the constriction. For simplicity we consider three possible orientations of the Ni tetrahedron (see Fig.~\ref{gfx:AC_geometric}c).

First we will discuss the magnetic configuration of the Ni particle as a function of orientation with respect to the leads. The magnetic configuration is characterized by the averaged magnetic moments of all four atoms in the tetrahedron.

The tetrahedra 2-2 and 1-1 are contacts, which have similar symmetry properties (see Fig.~\ref{gfx:AC_geometric}c). This is reflected in the electronic and magnetic properties, too. For instance, the distributions of the magnetic moments in the parallel configuration are symmetric with respect to the mirror plane. The averaged moments of the tetrahedra reach values of $0.70\,\mu_B$ for the 2-2 and $0.94\,\mu_B$ for the 1-1 orientation.

The domain wall in the anti-parallel configuration is located inside the tetrahedron for both orientations. The net magnetic moments are zero in both cases.

The 3-1 contact is asymmetric. The three atoms on the left and the one atom on the right hand side have different distances to the plateaus (see Tab.~\ref{tab:bonding_distances_particle}). The different distances between the atoms in the tetrahedron and the atoms in the plateaus ($P_L$, $P_R$) are the reason for an asymmetric distribution of the magnetic moments. In the parallel configuration, the average magnetic moment is $0.76\,\mu_B$, while in the anti-parallel configuration the net magnetic moment of the tetrahedron is $0.77\,\mu_B$. The domain wall is located between the left plateau ($P_L$) and the three atoms on the left hand side of the particle.

The calculated and spin-resolved conductances depending on the orientation of the tetrahedron are shown in Tab.~\ref{tab:Particle_g}. The calculated conductance values are of the same order of magnitude as for the constriction discussed above. A remarkable difference is that the conductance for minority states is larger than for the majority electrons in the parallel configuration which is in contradiction to the former cases of few atom contacts. However, this behavior of the spin-resolved conductances was observed for the single atom contacts with relative length variation $\Delta d/d_0<7.5~\%$ (cf. Fig.~\ref{gfx:SAC_g_all}b).

The conductance values in parallel and anti-parallel configuration are the highest for the 2-2 orientation of the particle. These relatively high values are caused by the number of bridging atoms per lead. Each contact is contacted by two atoms respectively. This implies a large ''effective'' contact area in comparison with the other orientations, where one atom of the tetrahedron is contacting the plateau(s) and symbolizes a ''bottleneck'' (see top in Fig.~\ref{gfx:AC_geometric}c).

The conductance in the anti-parallel case is nearly the same as for the parallel alignment of the lead magnetizations and the spin-resolved values for the conductance are degenerated. The degeneracy is caused by the symmetric position of the domain wall.

The conductance of the P configuration for the 2-2 orientation is similar to the conductance for orientation 1-1. The total conductance decreases from $2.25\,e^2/h$ to $2.19\,e^2/h$ and is influenced by the geometry. In orientation 2-2 and 1-1, the distances between the outermost atoms in the tetrahedron and the atoms in the plateau are nearly the same. The reduction of the number of bridging atoms from two to one can cause a blocking of the eigenchannels.

In the anti-parallel configuration the domain wall is located in the tetrahedron for the 2-2 and 1-1 orientation. As discussed above, the net moment of the particle is zero. The whole constriction has a mirror plane geometrically and a point symmetry magnetically. Therefore the spin-resolved conductance is degenerate. But the conductance values are reduced compared to the parallel configurations. For the 2-2 orientation the conductance is higher than for the 1-1 orientation which is again a result of the reduced number of bridging atoms.

By considering another orientation of the particle, one can end up with the 3-1 orientation. Now, the contact has no mirror plane. There are three atoms contacting the left plateau ($P_L$) and only one atom contacts the right plateau ($P_R$). The conductance in the parallel and antiparallel case is determined mainly by the increased distance between the left plateau and the left hand side of the particle (see Tab.~\ref{tab:bonding_distances_particle}). In Sec.~\ref{sec:contacts} we showed that tensile strain reduces the conductance which can be clearly seen by comparing the conductance values of the 2-2 and the 1-1 with the 3-1 orientation of the particle. In the AP configuration the conductance is slightly reduced compared to the parallel alignment of the lead magnetization due to the presence of a domain wall.

Furthermore, the threefold symmetry of the particle with respect to the $z$ axis does not fit the fourfold lead symmetry and causes blocking of the eigenchannels. This is reflected in the spin-resolved conductances. $g_{\uparrow\uparrow}$ is only slightly reduced since the spherical symmetry of the $s$-$p$ channel does not suffer from the symmetry misfit. $g_{\downarrow\downarrow}$ with pronounced $d$ character is, however, strongly perturbed by the symmetry misfit and is reduced with respect to the other orientations of the particle.

\begin{table}
\caption{\label{tab:Particle_MR} Calculated MR ratios in dependence on the orientation of the particle to the electrodes. The corresponding conductance values are listed in Tab.~\ref{tab:Particle_g}.}
\centering
\begin{ruledtabular}
\begin{tabular}{c|c|r}
\multicolumn{2}{c|}{orientation} 	& \multicolumn{1}{c}{MR ratio} \\
initial (parallel)	& final (anti-parallel)	& \multicolumn{1}{c}{[$\%$]} \\
\hline
 2-2 & 2-2 &  6.91 \\
 2-2 & 3-1 & 49.42 \\
 2-2 & 1-1 & 48.77 \\
\hline
 3-1 & 2-2 & -10.38 \\
 3-1 & 3-1 & 25.25 \\
 3-1 & 1-1 & 24.71 \\
\hline
 1-1 & 2-2 &  4.09 \\
 1-1 & 3-1 & 45.49 \\
 1-1 & 1-1 & 44.85 \\
\end{tabular}
\end{ruledtabular}
\end{table}

Assuming that the Ni particle can rotate freely triggered by an external magnetic field a variety of MR ratios can be obtained depending on initial and final orientation of the particle with respect to the leads. As mentioned above for simplicity we investigated three orientations of the particle which allows for exactly nine possible geometrical combinations for initial and final state of particles orientation (see Tab.~\ref{tab:Particle_MR}). The range of MR ratios spreads from +50 \% to -10 \%. The maximum values are in agreement with experimental results \cite{Viret02} but does not support the huge values reported elsewhere \cite{Garcia99,Garcia01,Chopra02,Sullivan05}. Interestingly the calculations predict a sign reversal of the MR ratio from positive to negative values depending on the particle orientation. Recently it was reported about reproducible large MR ratios ($+50~\%$ to $-20~\%$) at zero applied magnetic field in electromigrated permalloy nanocontacts (Ni$_{80}$Fe$_{20}$)\cite{Bieren2011}. Similar results were found for pure Ni nanocontacts.

\section{Conclusions}

We presented \textit{ab initio} calculations of the ballistic conductance for different Ni nanocontacts. It was shown that compressive strain increases the conductance whereas tensile strain reduces the conductance of a single-atom contact.

Furthermore, it was shown that the conductance of a few-atom contact strongly depends on the contact area as well as on the symmetry in the constriction. Reduction of contact area and of symmetry in the contact region with respect to the leads reduces the conductance.

For all cases we discuss the magnetoresistance and obtained values of up to 50 \% in agreement with experiment\cite{Agrait03,Viret02,Untiedt04}. The gigantic MR ratios\cite{Garcia99,Garcia00,Garcia01,Chopra02,Sullivan05} are not confirmed by our calculations. However, we predict a mechanism that can cause negative MR ratios. The mechanism is related to symmetry reduction in the contact region with respect to the leads and can be realized by a rotating Ni particle where the rotation is simulated by the external magnetic field.

\begin{acknowledgments}
This work was supported by the Deutsche Forschungsgemeinschaft DFG, Priority Program 1165: Nanowires and Nanotubes. One of the authors (S.A.) is a member of the International Max Planck Research School for Science and Technology of Nanostructures.
\end{acknowledgments}

\bibliography{refs}
\end{document}